\documentclass[]{elsarticle}

\usepackage{lineno,hyperref}
\usepackage{amsmath}
\modulolinenumbers[5]
\usepackage{subfigure}
\usepackage{xcolor}
\usepackage[margin=1in]{geometry}
\usepackage{siunitx}
\usepackage{adjustbox}

\journal{Journal}

\graphicspath{{figs/}}

\begin{document}

\begin{frontmatter}

\title{Transfer Functions of Proteinoid Microspheres}

\author{Panagiotis Mougkogiannis}
\author{Neil Phillips}
\author{Andrew Adamatzky}
\address{Unconventional Computing Laboratory, University of the West of England, Bristol, UK}

\cortext[cor1]{Corresponding author: Panagiotis.Mougkogiannis@uwe.ac.uk (Panagiotis Mougkogiannis)}

\begin{abstract}

\noindent Proteinoids, or thermal proteins, are inorganic entities formed by heating amino acids to their melting point and commencing polymerisation to form polymeric chains. Typically, their diameters range from \SIrange{1}{10}{\micro\metre}. Some amino acids incorporated into proteinoid chains are more hydrophobic than others, leading proteinoids to cluster together when they are present in aqueous solutions at specific concentrations, allowing them to grow into microspheres. The peculiar structure of proteinoids composed of linked amino acids endows them with unique properties, including action-potential like spiking of electrical potential. These unique properties make ensembles of proteinoid microspheres a promising substrate for designing future artificial brains and unconventional computing devices. To evaluate a potential of proteinoid microspheres for unconventional electronic devices we measure and analyse the data-transfer capacities of proteinoid microspheres. In experimental laboratory conditions we demonstrate that the transfer function of proteinoids microspheres is a nontrivial phenomenon, which might be due to the wide range of proteinoid shapes, sizes, and structures.


\end{abstract}


\end{frontmatter}


\section{Introduction}

Thermal proteins --- proteinoids~\cite{fox1992thermal} -- are produced by heating amino acids to their melting point and initiation of polymerisation to produce polymeric chains. The tri-functional amino acids undergo cyclisation at high temperature and become solvents and initiators of polymerisation for other amino acids~\cite{harada1958thermal,fox1992thermal}.   At moderate temperature in aqueous solution proteinoids swell into microspheres~\cite{fox1992thermal}. A communication between microspheres occurs via internal material exchange through the junctions~\cite{hsu1971conjugation}. A unique feature of proteinoid microspheres is that they maintain a steady state membrane potential 20--70~mV without any stimulating current and some microspheres in the population display the opposite polarization steadily~\cite{przybylski1985excitable}. Electrical membrane potentials, oscillations, and action potentials  are observed in the microspheres impaled with microelectrodes: the microspheres exhibit action-potential like spikes. The electrical activity of the microspheres also includes spontaneous bursts of electrical potential (flip-flops), and miniature potential activities at flopped phases~\cite{ishima1981electrical}. 
In \cite{adamatzky2021towards} we speculated that ensembles of proteinoid microspheres can be used to implement cascades of logical functions via spiking activity and thus form structures similar to micro-brains. This could pave the way for the development of programmable ``synthetic organisms,'' which could be used in fields as diverse as drug delivery and medical diagnostics. There may be far-reaching consequences for medicine and computing if synthetic molecular computers are developed and proteinoids are used to govern the behaviour of cells~\cite{adamatzky2016advances}. While considering proteinoid microspheres as novel organic nano-electronic circuits~\cite{adamatzky2021towards} a first step would be to investigate electrical characteristics of the substrate.

\section{Methods}
\label{methods}

The following materials were used in the experiments: L-glutamic acid (L-Glu)  CAS-No: 56-86-0, L-aspartic acid (L-Asp) CAS-No: 56-84-8, L-lysine (L-Lys) CAS-No: 57-82-49-2, L-histidine (L-His) CAS-No: 71-00-1, and L-phenylalanine (L-Phe) CAS-No: 63-91-2 purchased from Sigma Aldrich Ltd, UK,  with a reagent grade exceeding 98\%. Poly(L-lactic acid) (PLLA) with a molecular weight between 80,000 and 100,000 was purchased from Polysciences Ltd, UK. The method developed by Lugasi~\cite{lugasi2020designed} and Kolitz-Domb~\cite{kolitz2014engineered} was used as the foundation for the proteinoids' synthetic procedure.

The electrical properties of proteinoids was measured using an impedance -- amplitude -- phase frequency response network analyzer (C60, Cypher Instruments, London, UK~\cite{C60}). The C60 network analyzer applies \SI{2}{\volt}$_{pp}$ sinusoidal voltage waveform across the protenoids across the frequency range (\SI{10} {\hertz} to \SI{4} {\mega\hertz}). The network analyzer was connected to CypherGrapgh (V1.28.0) software package on a computer to control functionality and store measurements. The software evaluates the waveform after it passes through the sample and displays a Bode plot, the frequency response was analyzed and measurements stored. Using the C60, it is possible to determine how the proteinoids responds to an applied signal. Using the C60, changes in the proteinoid's electrical performance over time can also be recorded.

The electrical impedance of proteinoids was measured using a digital Inductance Capacitance Resistance (LCR) meter (model 891, BK Precision Ltd, UK~\cite{BK}). The LCR meter was configured to scan across the \SI{20} {\hertz} to \SI{300} {\kilo\hertz} frequency range applying \SI{1}{\volt}$_{rms}$ sinusoidal voltage waveform across the proteinoids. The phase angle describes the amount of time between the application of a voltage and the production of a current, whereas the impedance magnitude represents the opposition to current flow at a specific frequency. The magnitude and phase angle of an impedance can be used to calculate both its real and imaginary components. Due to its large scan range, the LCR metre model 891 is particularly helpful for measuring the electrical impedance of proteinoids.

The proteinoids were examined with FEI Quanta 650 Field Emission Scanning Electron Microscope (SEM). The FEI Quanta 650 is a highly effective tool for analysing the structure and composition of gold-coated material samples. This SEM can acquire high-resolution images of the sample's surface, allowing for a thorough analysis of its properties. The gold coating functions as both a barrier and a conductor, permitting the creation of the charged-particle beam crucial to the imaging capabilities of the SEM.

Different proteinoids solutions, including water and NaCl 0.15 M ionic solution, were deposited on a crystal of Nicolet iS 5 FTIR Spectrometer (Thermo Scientific). The spectra were collected throughout a scan range of 400 to 4000 $\mathrm{cm^{-1}}$ with a resolution of 4 $\mathrm{cm^{-1}}$. The Bicolet Omnic programme (OMNIC Series Software, Thermo Scientific) was used for data collection and spectrum analysis following FT-IR spectroscopy. Spectrum discrepancies between samples and NaCl 0.15 M ionic solution were identified by comparing them to a reference spectrum of the proteinoid sample in water.

\section{Results}

\begin{figure}[!tbp]
\centering
\includegraphics[width=0.99\textwidth]{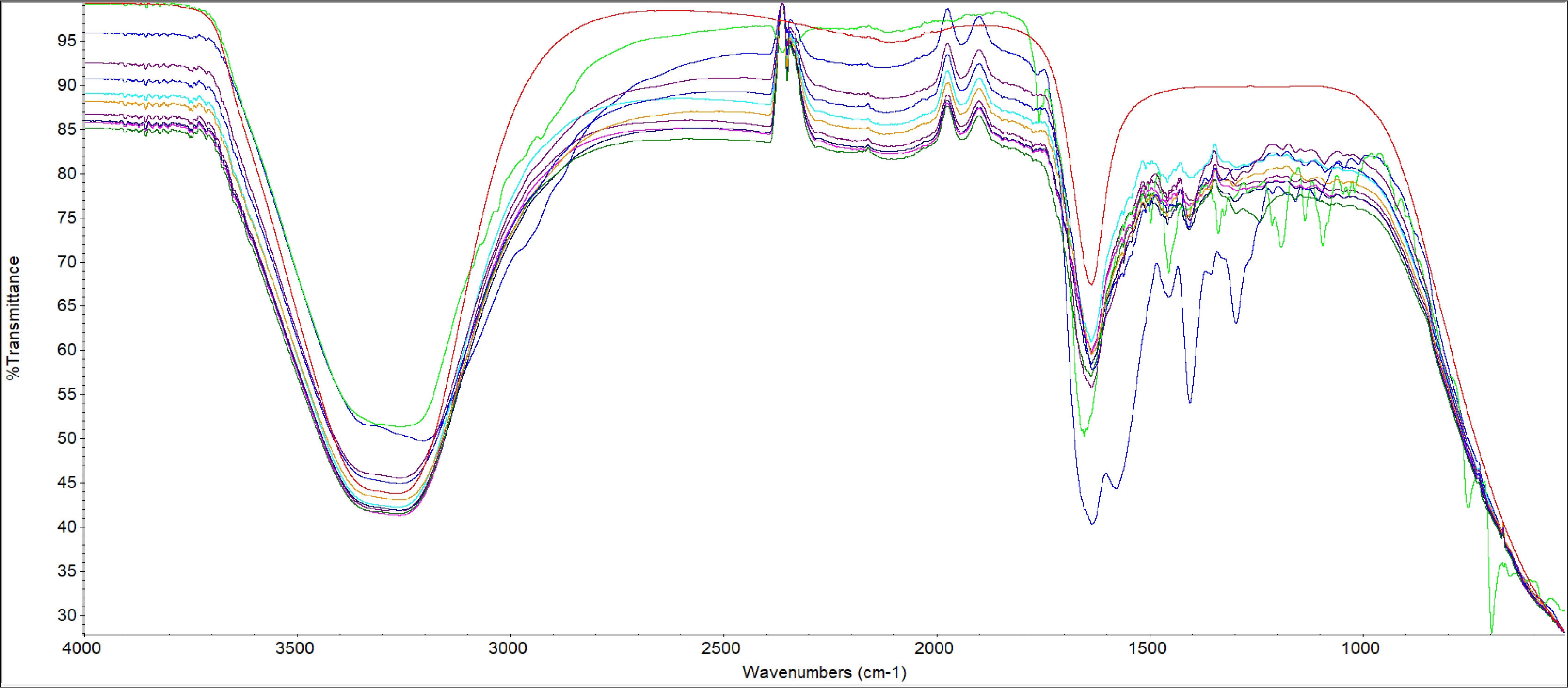}
\caption{FT-IR spectra of proteinoids with codes 12, 11, 6, 8, 12, 12F, 11 F, 10, 13, and 2 are displayed in the graphic, with the FT-IR vibrations of water shown in red. These spectra show that several chemical groups, such as amide I and II, aliphatic C-H stretching, and aromatic C-H stretching, are present in proteinoid solutions.
}
\label{p_11}
\end{figure}

Amide I and II, which are characteristic of the peptide backbone~\cite{dousseau1990determination}, are among the functional groups whose vibrations and absorptions are shown in the spectrum (Fig.\ref{p_11}). These peaks are attributable to the numerous functional groups found in amino acids, including the C-H stretching modes of methylene groups, the C-N and C-O stretching of amide bonds, and the C=O stretching of the carboxyl group. Several peaks become apparent in FT-IR analyses of the  proteinoid solutions (Fig.~\ref{p_11}). You may find these peaks at 1635, 1943, 2108, 2349, and 3258 $\mathrm{cm^{-1}}$. Amide II band, resulting to vibrations of peptide bond between amino acid residues in proteinoid, is represented by peak at 1635 $\mathrm{cm^{-1}}$. The stretching of the peptide group bonds causes the peak at 1943 $\mathrm{cm^{-1}}$, which corresponds to the amide I band. Supplemental FT-IR spectra of proteinoids provide additional evidence that proteinoids can be formed by heat condensation of amino acids, confirming the findings in the main article.

\begin{figure}[!tbp]
\centering
\includegraphics[width=0.8\textwidth]{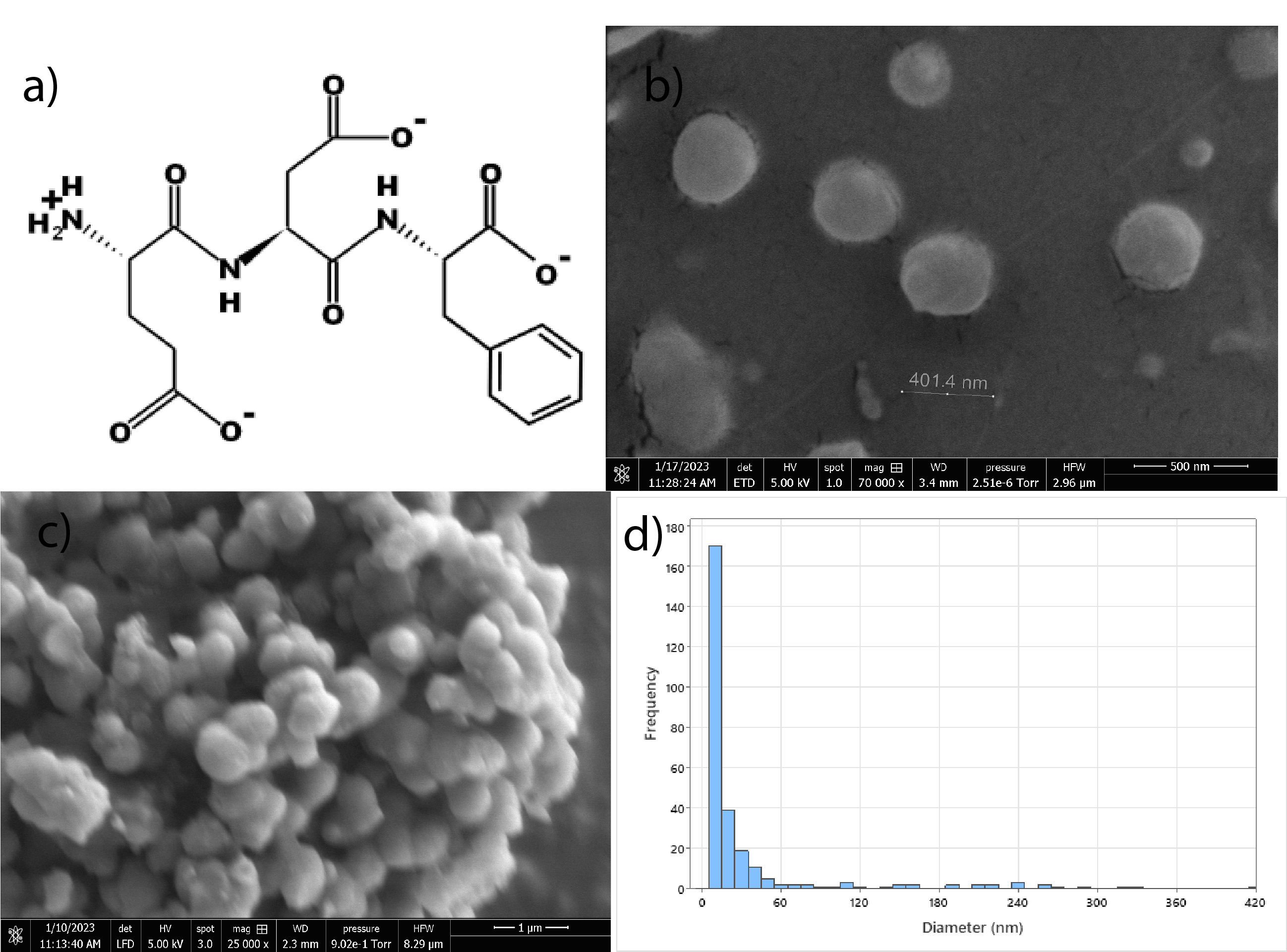}
\caption{(a)~Proteinoid L-Glu:L-Asp:L-Phe chemical structure. 
(b)~Examining the nanoscale structure of proteinoids nanospheres, with a bar representing 400 nm for scale. 
(c)~Proteinoid L-Glu:L-Asp:L-Phe , seen at a microscopic level, develop intricate structures that resemble the neural networks of a genuine brain. d) With a left-skewed distribution spanning from 2.5 nm to 420 nm, this histogram demonstrates the polydispersity of the L-Glu:L-Asp:L-Phe proteinoid nanospheres.
}
\label{p_1}
\end{figure}

The SEM images of proteinoid  microspheres, ranging in size from 2.5~nm to 400~nm (Figs.~\ref{p_1}b and \ref{p_1}c),  and the histogram (Fig.~\ref{p_1}d)  provide information about the structure and composition of proteinoids. The SEM images (Fig.~\ref{p_1}b and Fig.~\ref{p_1}c)  reveal the proteinoid's intricate structure. Larger aggregates of proteinoids lie with smaller, more irregularly shaped particles in the image. The image also shows the proteinoid's many grooves and bumps, which might be necessary for its electrical function. To learn more about how proteinoids are distributed in terms of size, we can look at the histogram of proteinoids whose diameters fall between 2.5 nm and 400 nm. The histogram displays a broad distribution, with the bulk of the proteinoids lying between the 5~nm and 400~nm size boundaries. This suggests that the proteinoids consist of a wide range of constituent building blocks, each of which is of a different size.  



\begin{figure}[!tbp]
\centering
\includegraphics[width=0.8\textwidth]{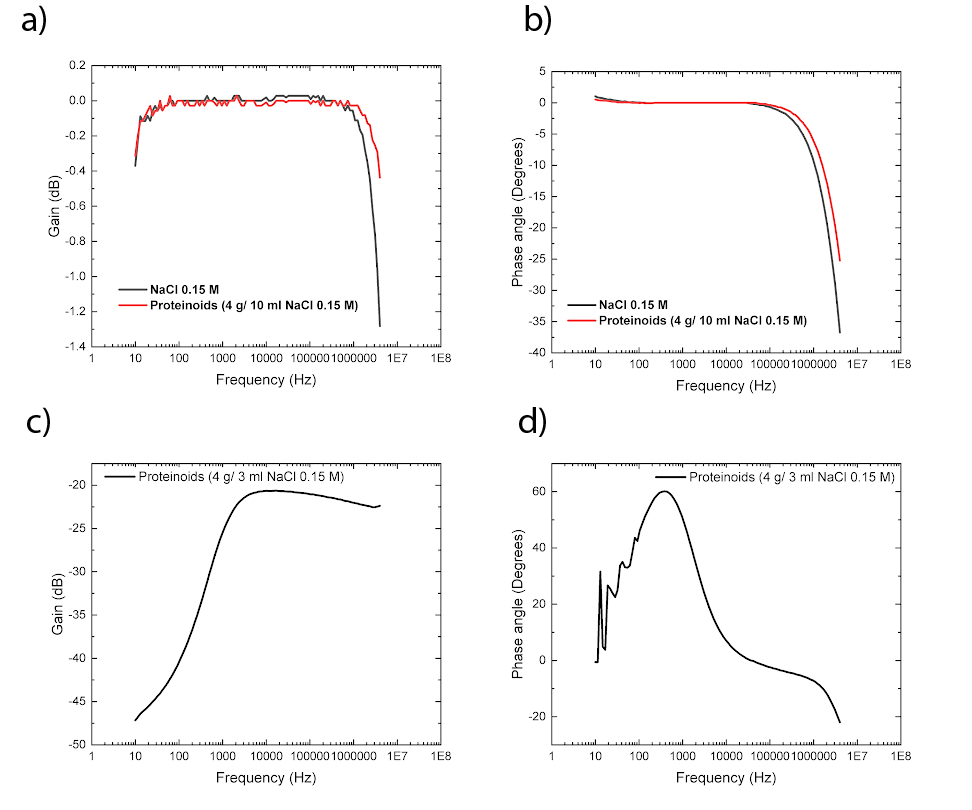}
\caption{a)The Bode plot of a proteinoid L-Glu:L-Asp solution in 0.15 M NaCl shows the gain response of the proteinoid to changing frequencies indicating the stability of the system.
b) The Bode plot of a proteinoid L-Glu:L-Asp solution in 0.15 M NaCl reveals significant differences in the magnitude and phase of the response, indicating a strong influence of the salt concentration on the proteinoid’s structure and function.
c) Bode plot of proteinoids L-Glu:L-Asp solution (5g /3ml NaCl 0.15 M) for gain (dB) d) Bode plot of proteinoids L-Glu:L-Asp solution (5g /3ml NaCl 0.15 M) for phase angle.}
\label{p1}
\end{figure}

Measuring the proteinoid solution NaCl 0.15 M with a C60 impedance-amplitude analyzer over a frequency range of 10~Hz to 
4~MHz yields a Bode plot that provides insight into the 
substrate's characteristics. By depicting the signal's amplitude and phase as a function of frequency, this graph can be used to verify the system's stability.

The signal measured by the C60 impedance-amplitude analyzer in a proteinoid solution of NaCl 0.15 M is shown in the Bode plot in Fig.~\ref{p1}a. The graph displays the signal strength as a function of frequency. As can be seen in the plot, the signal strength rapidly improves up to about 100~Hz, after which it flattens out. From this, we can infer that the signal is robust up to a frequency of $10^{6}$~Hz, but starts to degrade above that point.

Phase as a function of frequency is depicted in Fig.~\ref{p1}b. In this diagram, we can see that the phase of the signal is fairly constant up to about $\mathrm{10^{5}}$~Hz, and then it starts to drop off as the frequency increases. The signal acts in a predictable way up to 500~Hz, but after that it gets more and more unpredictable.
Impedance measurements could be productive in discovering properties of proteinoids solutions, including resistance, capacitance, and inductance that they possess. Frequency dependent attenuation describes how a signal weakens as it passes through a medium based on its frequency. Dissolution of macromolecules in solution and the formation of hydrogen bonds between amino acids contribute to the breakdown of proteinoids. Higher attenuation occurs at lower frequencies and vice versa.
The results reveal that up to a certain signal frequency (3088 Hz), attenuation increases and then decreases. This means that at higher frequencies, attenuation will be stronger because hydrogen bonds between amino acids will occur more frequently. The degree of attenuation is dependent not only on frequency, but also on solution NaCl concentration and proteinoid concentration.Increasing the NaCl concentration results in a greater attenuation, while increasing the proteinoid concentration has the opposite effect (see Fig.~\ref{p1}).

Figure~\ref{p1}c shows a negative gain  from 10~Hz to $\mathrm{10^7}$~Hz with a maximum gain of ~20~dB at 1000~Hz. The phase angle is positive and increases with increasing frequency, reaching a maximum 60$^{\circ}$ at 1000~Hz. Proteinoid transfer function in a 0.15~M NaCl solution is a non-trivial and frequency-dependent phenomenon. The composition of proteinoid, specifically L-Glu:L-Asp, is crucial to the transfer function and determines the signal attenuation.

Transfer function of proteinoid in a 0.15~M NaCl solution is determined by the collective properties of amino acids and their interactions with solution ions. Proteinoids are large molecules composed of subunits held together by covalent bonds, such as amino acids. Covalent bonds in a solution can provide energy to the ions through interactions. The composition of the proteinoid, specifically L-Glu:L-Asp, is crucial to the proteinoid's transfer function. L-Glu:L-Asp is a dipeptide composed of two amino acids: glutamic acid and aspartic acid. This dipeptide is important in the transfer function of proteinoid because it can form hydrogen bonds with the ions in the solution. This causes a greater transfer of energy, which results in a greater signal attenuation.
The transfer function of proteinoids is likewise heavily influenced by the signal's frequency. A greater attenuation of the signal occurs at lower frequencies because the proteinoids are more mobile in the solution and interact more strongly with the ions. Proteinoids are restricted in their mobility at higher frequencies, resulting in less attenuation of the signal~Fig.~\ref{p1}d.

\begin{table}[!tbp]
\centering
 \caption{ Resistance, Impedance, and Capacitance of Proteinoids at 300kHz.}
 \label{table:1}
 \begin{tabular}{||c c c c c ||} 
 \hline
   Proteinoid  & \textbf{Resistance } (k$\Omega$) & \textbf{Impedance } (k$\Omega$)  & \textbf{Capacitance (nF)} & \textbf{Code of Exp.} \\ [0.5ex] 
 \hline\hline
 \textbf{L-Glu:L-Asp} & 0.4839 & 0.4833
& 64.82 & 2  \\ 
 \textbf{L-Glu:L-Phe:L-His} & 0.09754 & 0.09732
& 434.9 & 8  \\
 \textbf{L-Lys:L-Phe:L-His} & 1.679 & 1.683&3.327 & 13 \\
 \textbf{L-Glu:L-Phe} & 0.04378 & 0.05152&29.85 &10   \\
 \textbf{L-Glu:L-Asp:L-Lys} & 0.4706 & 0.4805 &35.23 & 11 \\
 \textbf{L-Glu} & 0.2456 & 0.2471&400.2 &12 \\
 \textbf{L-Glu: L-Phe:PLLA} & 0.2318 & 0.2301  & 170.2 & 11F \\
 \textbf{L-Lys:L-Phe-L-His:PLLA} & 0.04574 & 0.04755 & -656.6 & 12F \\
 \textbf{L-Glu:L-Asp:L-Phe} & 0.3756 & 0.3736 & 42.23  & 6 \\ [1ex] 
 \hline
 \end{tabular}
\end{table}

The electrical properties of proteinoids are detailed in Tab.~\ref{table:1}. 
Data analysis reveals the proteinoids' resistance (in Ohms), impedance (in Ohms), and capacitance (in nF).


\begin{figure}[!tbp]
\centering
\includegraphics[width=0.8\textwidth]{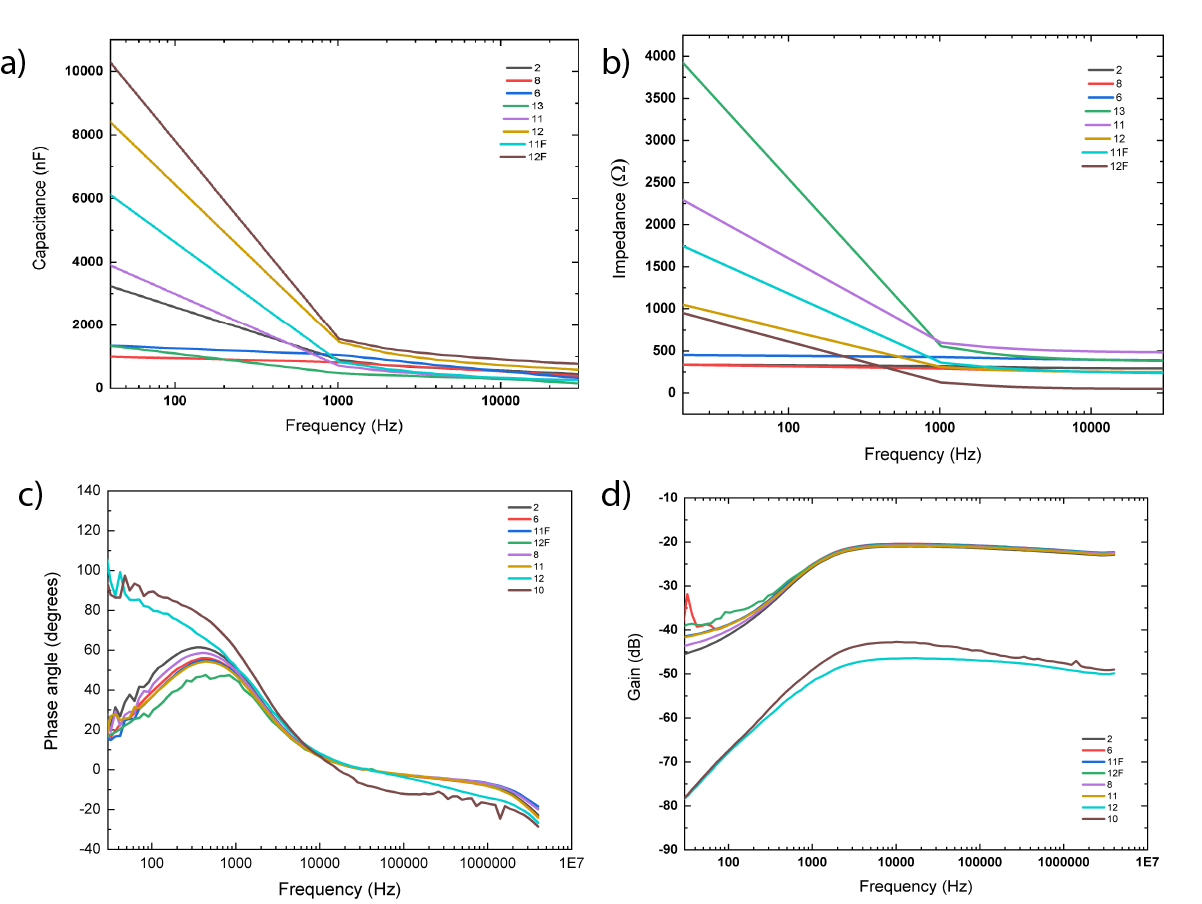}
\caption{a) This wide variation in capacitance values between proteinoids is indicative of the structural and electrical complexity of these molecules. b) The graph shows how the impedance of proteinoids changes at different frequencies, which shows how these biomolecules behave electrically in a complicated way. c) The figure shows the phase angle of the proteinoids in relation to their frequency. d) Gain, measured in decibels (dB), of various proteinoids at a range of low to high Hz frequencies. Each proteinoid clearly exhibits a different response over the frequency range, making them each suitable for their own specific uses.
}
\label{p6}
\end{figure}

In recent years, scientists have explored the intriguing effect of negative capacitance (proteinoid code 12F) in Table~\ref{p6} \cite{khan2015negative,ershov1998negative,zhirnov2008negative,iniguez2019ferroelectric}. When the voltage across a component is increased, the energy stored in the electric field decreases, a phenomenon known as negative capacitance. This only happens very seldom in some systems, such as carbon nanotubes and graphene in field effect transistors~\cite{srimani2017negative,tamersit2022leveraging}. The efficiency and effectiveness of electrical components can be enhanced by harnessing this effect. According to Johnscher's work~\cite{jonscher1986physical}, a time domain analysis of negative capacitance under a step function bias is the proper way to investigate this phenomenon. 

The quantity of a charge stored in the proteinoids at various frequencies is depicted in the Fig.~\ref{p6}a of capacitance 
versus frequency.  Impedance versus frequency graph (Fig.~\ref{p6}b) illustrates the frequency-dependent resistance to current flow. The Figs.~\ref{p6}a and~\ref{p6}b illustrate crucial details of the proteinoids' electrical properties. Each parameter, capacitance and impedance, responds differently to changes in oscillation frequency. Both capacitance and impedance are lowered as frequency rises and raised as frequency falls. This indicates that at low frequencies the proteinoids have greater charge storage capacity and current resistance.

The phase angle of a Bode plot is a measurement of the phase shift as a function of frequency between the input and output signals from a system. It demonstrates how the proteinoid signal's amplitude and phase fluctuate in relation to the input signal's frequency. 
The phase angle of the proteinoids as a function of frequency is shown in the Fig~\ref{p6}c. The phase angles of proteinoids with codes 6 and 12 differ significantly from those of other proteinoids. The interaction between the proteinoids and other molecules is impacted by this variation in phase angle.


\section{Discussion}

The correlation between the concentration of proteinoids in a solution and its electrical conductivity has been investigated. In laboratory experiments, we demonstrated that the electrical transfer function of ensembles of proteinoid microspheres is nontrivial and sensitive to the exact composition of the proteinoids and the solution in which the microspheres reside.

Is there a correlation between the amount of proteinoids in a solution and how well it conducts electricity?

The correlation between proteinoid concentration and signal attenuation is depicted in Fig.~\ref{p1}. Signal strength is reduced proportionally to the number of microspheres present. This data demonstrates that the concentration of proteinoids in an electrolyte solution influences the electrical properties of polypeptides, such as their ability to transmit electrical signals.


The ability of thermal proteins to both accept and resist electrical signals, as well as the number of charges carried by amino acids, define their distinctive electrical properties. The electrical properties of proteinoids are determined by the total number of amino acids in a cell, which in turn affects the number of charges carried by the molecules. The ability of thermal protein molecules to accept or resist electrical signals increases as the concentration of proteinoids and number of charges increase. Consequently, more energy is required to transmit a signal, which increases the attenuation.

\begin{figure}[!tbp]
\centering
\includegraphics[width=0.7\textwidth]{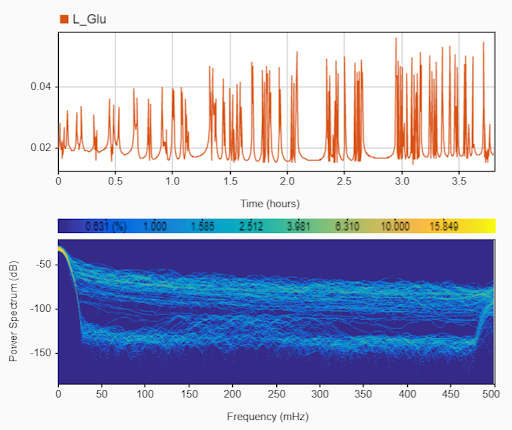}
\caption{Using a PicoLog data logger and a platinum-iridium electrode, we can observe the dynamic nature of the proteinoid L-Glu by tracking its electrical oscillations. 
}
\label{ppp9992}
\end{figure}

Employing proteinoids in biohybrid computing systems requires accurate measurements of proteinoid-environment interactions. Therefore, signal transfer analysis of proteinoids is a useful technique, as it enables precise determination of signal strengths and durations between proteinoids and their surroundings. Signal transfer analysis can be utilised to discover the dynamics of a system, such as the interactions between proteinoids or between proteinoids and their surroundings. This knowledge may enable complex algorithms to be programmed into proteinoid, giving them the ability to adapt to new situations. Based on the provided data, proteinoids might be programmed to respond to molecules as a switches or logic gates. Proteinoids' electrical spikes (Fig.~\ref{ppp9992}) are a type of protein-based circuitry that might be used to monitor and operate mining logical circuits, as well as for other computational purposes. Proteinoids' amino acids and other building blocks might be connected to form a biological network. Proteinoids produce electrical spikes when the electrical fields generated by the circuit's components interact. This interaction produces a voltage spike that might be used to monitor and control the operation of mining logic circuits.

The frequency-dependent attenuation of proteinoids is thought to be connected to their structure and composition, despite not being fully understood. Within proteinoids, amino acids of different polarity are organised in a particular pattern. It is believed that dissimilar polarities dampen and absorb electromagnetic and acoustic waves of specific frequencies. Due to their effect on the electrostatic properties of the molecule as a whole, polarizability, hydrophobicity, and ionic charge all contribute to proteinoids' electrical properties. When proteinoids are in an electric field, polarizable molecules experience a stronger force than nonpolarizable molecules. This is important for proteinoids because it influences how the molecules interact with one another and, ultimately, how the proteinoid behaves electrically. Polarizable molecules, for instance, can form hydrogen bonds, which can help stabilise the structure of a proteinoid and improve its electrical properties. In the case of microspheres, the hydrophobic portions of the molecule interact with water molecules to generate an electrical charge. This charge will alter the proteinoids' electrical properties, making it easier for them to interact with their surroundings. For example, a negatively charged proteinoid may interact more strongly with a positively charged environment.


\begin{figure}[!tbp]
\centering
\includegraphics[width=0.6\textwidth]{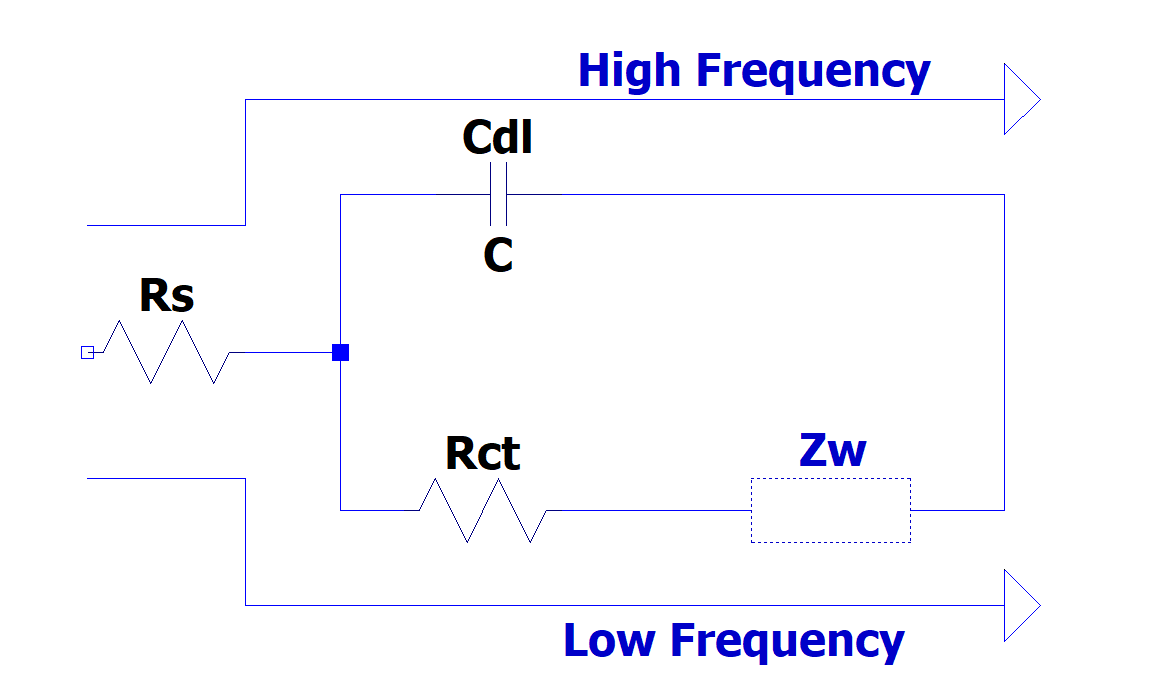}
\caption{The figure shows the electrical flow in the Randles cell, with resistance $\mathrm{R_{s}}$ due to the solution, $\mathrm{R_{ct}}$  as the charge transfer resistance, $\mathrm{C_{dl}}$ as the double layer capacitance, and $\mathrm{Z_{w}}$ as the impedance. The figure illustrates the current flow at both high and low frequencies.
}
\label{ppp6}
\end{figure}


Randles cells (Fig.~\ref{ppp6}), a specific form of electrochemical cell, are utilised extensively in the field of electrochemistry. John Edward Brough Randles, a British scientist, conceived of it in 1947~\cite{randles1947kinetics}. Two-electrode systems with a working electrode and a counter electrode immersed in a reactant solution constitute Randles cells. It has many applications in electrochemical research, such as evaluating the current-voltage (I-V) properties of a material or determining the activity of a substance in a solution.


A Randles cell contains two electrodes and generates electricity by applying a voltage difference between them. In a chemical reaction, the electrode that contacts the reactant solution is known as the working electrode, while the electrode that serves as a reference point for the potential difference between the working electrode and the current electrode is known as the counter electrode. The reactants in the solution undergo a chemical reaction at the working electrode due to the potential difference. It is possible to use the results of such a study to evaluate the amplitude of the current.


The electrical properties of proteinoids are essential to the Randles cell because they allow an electric current to flow through the cell when a voltage is applied across its terminals. Applications of the Randles cell include research on the effects of electric fields on proteinoids and the effect of pH on the electrical properties of proteins, as well as the advancement of fuel cell technology~\cite{laribi2016impedance,fouquet2006model}.


The Nyquist plot is a helpful tool for understanding the electrochemical properties of proteinoids. By analysing a molecule's Nyquist plot, which displays the real and imaginary components of a proteinoid's electrical properties separately, it is possible to learn more about its response to electrical stimulation. This can shed light on how a proteinoid might react to various environmental conditions or interact with other molecules. The electrical properties of proteinoids can be better comprehended with the aid of a Nyquist plot, which depicts the propagation of electrical signals through a molecule.

\begin{figure}[!tbp]
\centering
\includegraphics[width=1\textwidth]{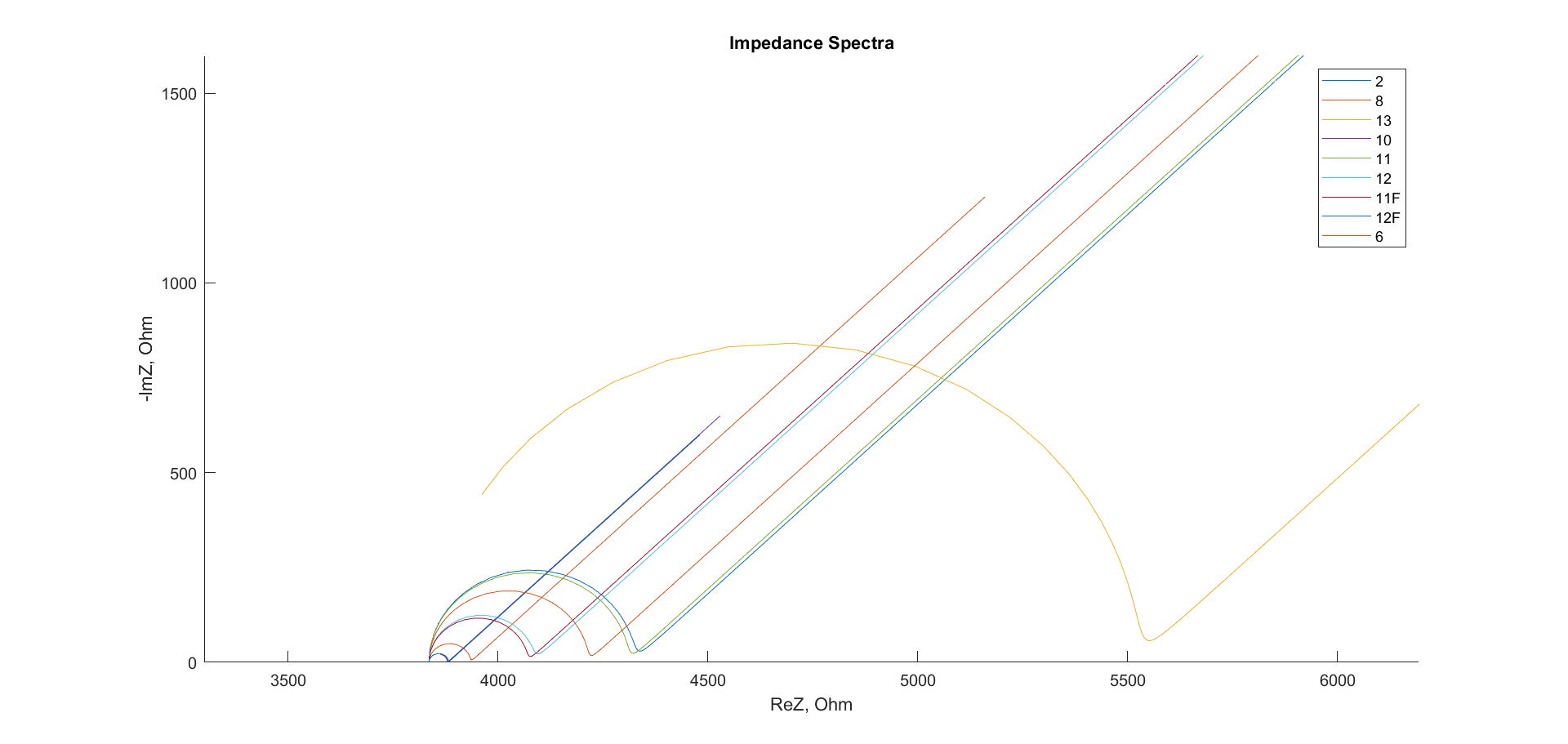}
\caption{Proteinoids with the following codes: 2,8,13,10,11,12,11F,12F,6 as plotted on a Nyquist diagram. The water resistance is 3836 Ohm, and the double layer capacitance, charge transfer resistance, and impedance values are all taken from Table \ref{table:1}.
}
\label{ppp999}
\end{figure}


The Nyquist plot (Fig.~\ref{ppp999}) illustrates the complex impedance of a system as a function of frequency. It is a two-dimensional graph depicting the real and imaginary components of the system's impedance. The imaginary portion of the impedance response is plotted vertically, while the real portion is plotted horizontally. The Nyquist plot demonstrates the connection between frequency and impedance. At low frequencies, the Nyquist plot reveals the system's resistance due to diffusion. At higher frequencies, the Nyquist plot reveals the system's solution's resistance. The Nyquist plot is a crucial tool for analysing the electrical behaviour of systems. It is used to measure the frequency response and impedance of a system, as well as to determine the stability of a system. It is also used to determine the capacitance and inductance effects of a system. Additionally, it can be used to determine the effects of parasitic components, such as inductance and capacitance, on a system.


In a Nyquist plot, the diameter of the semicircle represents the magnitude of the open-loop transfer function. The magnitude of the transfer function, which is the output-to-input ratio, represents the degree to which an input signal is amplified or dampened.


A larger L-Lys:L-Phe:L-His semicircle, as shown in Fig.~\ref{ppp9992}, is indicative of a stronger proteinoid-proteinoid interaction. This is because a system with a larger diameter is more sensitive to higher frequencies and can therefore interact with other proteins and molecules more readily. Consequently, L-Lys:L-Phe:L-His proteinoids are more likely to form stronger bonds with other proteins. In contrast, the L-Glu proteinoid is not as effective at binding with other proteinoids due to its smaller semicircle diameter, which indicates that it is less receptive to higher frequencies.

Understanding the electrical spikes caused by proteinoids will be a significant focus of future research. How the electrical oscillations of L-Glu  (Fig.~\ref{ppp9992}) relate to the proteinoid's physiological roles requires additional examination. Future experiments will examine the effect of varying proteinoid concentrations on electrical oscillations and their potential relevance to other biological processes. Scientists hope to gain a greater understanding of how proteinoids function in cellular communication and whether or not they can be modified to produce specific effects.




\section{Funding} 
The research was supported by EPSRC Grant EP/W010887/1 ``Computing with proteinoids''.The funders played no role in the design of the study and collection, analysis, and interpretation of data.

\section{Acknowledgements}
Authors are grateful to David Paton for helping with SEM imaging.


\end{document}


\begin{frontmatter}

\title{Electronic Supplementary Information (ESI) \\ Transfer Function of Proteinoid Microspheres}



\author{Panagiotis Mougkogiannis}
\author{Neil Phillips}
\author{Andrew Adamatzky}
\address{Unconventional Computing Laboratory, University of the West of England, Bristol, UK}

\cortext[cor1]{Corresponding author: Panagiotis.Mougkogiannis@uwe.ac.uk (Panagiotis Mougkogiannis)}

\end{frontmatter}

\begin{figure}[h]
\centering
\includegraphics[width=0.8\textwidth]{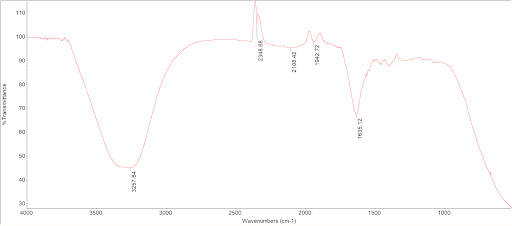}
\caption{Strong chemical connection between L-Glu, L-Asp, and L-Phe is shown by several peaks in the FT-IR spectrum.
}
\label{p_s1}
\end{figure}
\begin{figure}[h]
\centering
\includegraphics[width=0.8\textwidth]{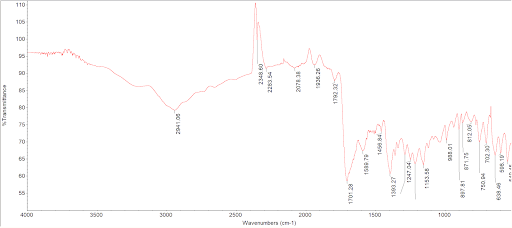}
\caption{FT-IR spectrum of the solid L-Glu:L-Asp:L-Phe proteinoid showing the presence of amide I and II stretching bands, indicating the presence of structure elements in the proteinoid.
}
\label{p_s2}
\end{figure}

More FT-IR peaks may be observed between 500 and 1590 $\mathrm{cm^{-1}}$ in the solid proteinoid L-Glu:L-Asp:L-Phe~(Fig.~\ref{p_s2}) compared to the liquid proteinoid L-Glu:L-Asp:L-Phe  (Fig.~\ref{p_s1}), probably as a result of structural differences between the two states. The proteinoid's L-Glu:L-Asp:L-Phe constituents crystallise into a rigid, three-dimensional structure. Since more angles and bonds can be formed in the solid state, the FT-IR spectrum of the solid demonstrates a greater number of peaks between 500 and 1590 $\mathrm{cm^{-1}}$ than the liquid. During the liquid phase of the proteinoid, the L-Glu:L-Asp:L-Phe components are disorganised and free to move. In the range from 500 to 1590 $\mathrm{cm^{-1}}$, there are fewer FT-IR peaks, indicating that these components do not form the same angles and bonds as they do in the solid phase. Because molecular vibrations in the liquid phase are more disorganised and unpredictable than in the solid phase, there are fewer FT-IR peaks.

\begin{figure}[h]
\centering
\includegraphics[width=0.8\textwidth]{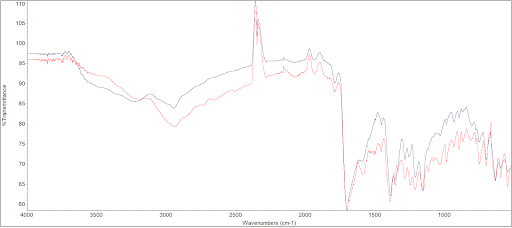}
\caption{ FT-IR spectra of Glu-Asp-Phe showing the differences between drying with heating (red) and lyophilisation (blue).
}
\label{p_s3}
\end{figure}

The FT-IR spectrum of the proteinoid L-Glu:L-Asp:L-Phe is unaffected by drying (Fig.~\ref{p_s3}). This has been demonstrated by two types of drying, namely heating and lyophilization. However, the secondary, tertiary, and quaternary structures of the proteinoid may change when the sample is heated to \SI{50} {\celsius}. In contrast, lyophilization is a milder drying process that involves freezing a sample before vacuuming away any remaining moisture. This typically reduces the size of proteinoids, but has no notcable effect on FT-IR. Furthermore, L-Glu:L-Asp:L-Phe FT-IR is unaffected by the drying process. This is the case because the FT-IR measures molecular vibrations, which are relatively unaffected by drying. Regardless of the method used to dry a substance, the FT-IR spectrum will display the same molecular vibrational strength and frequency.

\begin{figure}[h]
\centering
\includegraphics[width=0.4\textwidth]{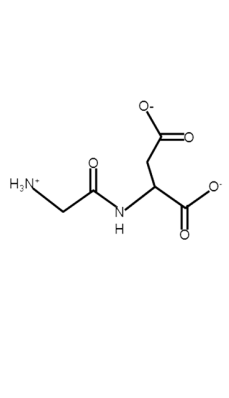}
\caption{The figure shows the chemical structure of the dipeptide L-Glu:L-Asp, which is composed of the two amino acids L-Glutamic acid and L-Aspartic acid. }
\label{p_ss1}
\end{figure}

\begin{figure}[h]
\centering
\includegraphics[width=0.8\textwidth]{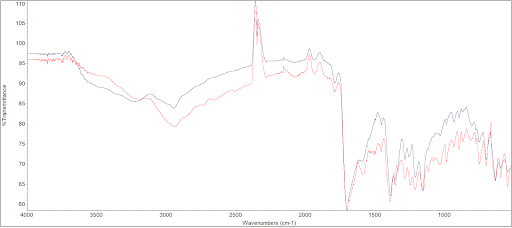}
\caption{ An FT-IR spectrum of L-Glu:L-Asp is displayed in the figure. The stretching of the C=O bond and the C-O bond can be seen as two different peaks in the spectra, located at 1703 $\mathrm{cm^{-1}}$ and 1389 $\mathrm{cm^{-1}}$, respectively. This proves the Glu:Asp sample is made up of both amino acids.}
\label{p_s5}
\end{figure}

The L-Glu:L-Asp proteinoid~(Fig.~\ref{p_ss1}) exhibits absorption peaks characteristic of amides in the range of 1650–1750 $\mathrm{cm^{-1}}$. This illustration (Fig.~\ref{p_s5}) shows how FT-IR spectroscopy can be used to effectively characterise proteins and peptides.

The 635 $\mathrm{cm^{-1}}$ peak is attributable to the C-H stretching modes of the methylene groups in the side chains of glutamic and aspartic acids. The peak at 700 $\mathrm{cm^{-1}}$ is due to the C-H stretching of the side chain of aspartic acid. The 934 $\mathrm{cm^{-1}}$ peak is due to the C-H out-of-plane bending of the aspartic acid side chain. The peak at 1161 $\mathrm{cm^{-1}}$ is due to the C-H in-plane bending of the side chain of aspartic acid. The 1210 $\mathrm{cm^{-1}}$ peak is caused by the C-C stretching of amide bonds in glutamic and aspartic acid. The peak at 1287 $\mathrm{cm^{-1}}$ is attributable to the C-N stretching of amide group bonds in glutamic and aspartic acid. The 1704 $\mathrm{cm^{-1}}$ peak is due to the stretching of the C=O bond in the carboxyl group of both glutamic and aspartic acid. The peak at 1791 $\mathrm{cm^{-1}}$ is caused by the stretching of amide bonds in glutamic and aspartic acid. This peak is caused by the stretching of the C=O bond in the carboxyl group of glutamic and aspartic acids. The peak at 2950 $\mathrm{cm^{-1}}$ in the spectrum of glutamic acid is caused by the O-H stretching of the carboxyl group. The peak at 3232 $\mathrm{cm^{-1}}$ is caused by the stretching of amide bonds in both glutamic and aspartic acid. In conclusion, the FT-IR spectra of L-Glu:L-Asp display a number of distinct peaks that shed light on its structure and properties. These peaks are due to the numerous functional groups present in amino acids, such as the C-H stretching modes of methylene groups, the C-N and C-O stretching of amide bonds, and the C=O stretching of the carboxyl group.

\begin{figure}[h]
\centering
\includegraphics[width=0.8\textwidth]{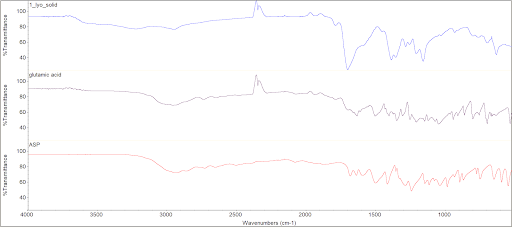}
\caption{ The figure shows the Fourier Transform Infrared (FT-IR) spectrum of a L-Glu:L-Asp proteinoid, which is composed of the two amino acids L-glutamic acid and L-aspartic acid.}
\label{p_s9}
\end{figure}
